\documentclass[letterpaper,twocolumn,10pt]{article}
\usepackage{usenix}
\usepackage[subtle]{savetrees}

\usepackage{times}
\usepackage[T1]{fontenc}
\usepackage[utf8]{inputenc}
\usepackage[english]{babel}
\usepackage[hyperref]{backref}
\usepackage{outlines}
\usepackage{graphicx,xcolor,xspace}
\usepackage{fancyvrb}
\usepackage[newfloat=true,frozencache]{minted}
\setminted[rust]{fontsize=\footnotesize}
\usepackage{booktabs}
\usepackage{tabu}
\usepackage{tabularx}
\usepackage[skip=-5pt]{subcaption}
\usepackage{paralist}
\usepackage[hang,flushmargin]{footmisc}

\usepackage[htt]{hyphenat}
\usepackage{ifthen}

\usepackage[compact,small]{titlesec}

\usepackage{enumitem}
\usepackage{caption}
\usepackage{breakurl}
\usepackage{etoolbox}
\appto\UrlBreaks{\do\-}
\makeatletter
\def\maxwidth{ %
  \ifdim\Gin@nat@width>\linewidth
    \linewidth
  \else
    \Gin@nat@width
  \fi
}
\makeatother

\makeatletter
{\catcode`\!=8 
 \catcode`\Q=3 
\long\gdef\given#1{88\fi\Ifbl@nk#1QQQ\empty!}
\long\gdef\blank#1{88\fi\Ifbl@nk#1QQ..!}
\long\gdef\nil#1{\IfN@Ught#1* {#1}!}
\long\gdef\IfN@Ught#1 #2!{\blank{#2}}
\long\gdef\Ifbl@nk#1#2Q#3!{\ifx#3}
}
\makeatother
\def\expblank{\expandafter\blank\expandafter} 

\definecolor{fgcolor}{rgb}{0.345, 0.345, 0.345}

\usepackage{framed}
\makeatletter
 {\par\unskip\endMakeFramed%
 \at@end@of@kframe}
\makeatother

\definecolor{shadecolor}{rgb}{.97, .97, .97}
\definecolor{messagecolor}{rgb}{0, 0, 0}
\definecolor{warningcolor}{rgb}{1, 0, 1}
\definecolor{errorcolor}{rgb}{1, 0, 0}
\newenvironment{knitrout}{}{} 
\usepackage{alltt}
\newcommand{\cut}[1]{}
\newcommand{\paragrapha}[1]{\vspace{0.02in}\noindent{\bf #1}}

\usepackage[textsize=tiny,textwidth=0.6in]{todonotes}
\newcommand{\allnotes}[1]{}
\renewcommand{\allnotes}[1]{\textit{#1}}

\if\expblank{\allnotes{a}}%

\else

\fi

\def\ie{{i.e.},\xspace}
\def\eg{{e.g.},\xspace}

\newcommand{\name}{Bertha\xspace} 
\newcommand{\tunnel}{Chunnel\xspace}
\newcommand{\tunnels}{Chunnels\xspace}

\newcommand{\eat}[1]{}
\date{}
\hypersetup{pdfstartview=FitH,pdfpagelayout=SinglePage}

\usepackage{fdsymbol}
\newboolean{anonymous}
\setboolean{anonymous}{false}
\usepackage[normalem]{ulem}

\begin{document}
\title{Bringing Reconfigurability to the Network Stack}
\author{Akshay Narayan$^{\varheartsuit}$, Aurojit Panda$^{\spadesuit}$, Mohammad Alizadeh$^{\varheartsuit}$,\\
Hari Balakrishnan$^{\varheartsuit}$, Arvind Krishnamurthy$^{\clubsuit}$, Scott Shenker$^{\vardiamondsuit}$\\
\textrm{$^{\varheartsuit}$ MIT CSAIL, $^{\spadesuit}$ NYU, $^{\clubsuit}$ University of Washington, $^{\vardiamondsuit}$ UC Berkeley and ICSI}}
\pagestyle{plain}
\maketitle
\begin{abstract}
\vspace{2pt}
Reconfiguring the network stack allows applications to specialize the implementations of communication libraries depending on where they run, the requests they serve, and the performance they need to provide. Specializing applications in this way is challenging because developers need to choose the libraries they use when writing a program and cannot easily change them at runtime. This paper introduces \name, which allows these choices to be changed at runtime without limiting developer flexibility in the choice of network and communication functions.
\name allows applications to safely use optimized communication primitives (including ones with deployment limitations) without limiting deployability. Our evaluation shows cases where this results in $16\times$ higher throughput and $63\%$ lower latency than current portable approaches while imposing minimal overheads when compared to a hand-optimized versions that use deployment-specific communication primitives.
\end{abstract}
\eat{
Most modern applications access the network using communication libraries rather than traditional network APIs. This is because the current network stack is not extensible and hence cannot support the rich variety of connections, including ones that encrypt or serialize data, modern applications use. However, most communication libraries are not designed to compose with other libraries, and this limits their use.
Additionally, many communication libraries can only be used in specific deployments, and the use of these libraries impedes portability. In this paper, we argue that to resolve these shortcomings, we need an extensible and runtime-reconfigurable network stack. We propose a design for such a stack and prototype our design in a system named \name. We demonstrate the generality and benefit of our design using three case studies.
}

\sloppy
\section{Introduction}\label{s:intro}
Distributed and cloud computing applications require the endpoint network stack to provide more than just (reliable or unreliable) communication channels accessed via the sockets API.
Many applications use the network to implement higher-level communication primitives such as remote procedure calls (RPCs)~\cite{birrell1984implementing} and publish-subscribe (pub/sub)~\cite{virtual-synchrony} communication. 
To perform well and meet their service-level objectives (SLOs), applications need low latency and/or high throughput from these libraries. 

Operating systems and shared libraries have evolved to make it simpler to write networked applications that implement complex logic while meeting stringent performance requirements.
For example, new kernel network interfaces such as io\_uring~\cite{iouring} and kernel bypass libraries such as DPDK~\cite{dpdk} reduce software overheads and improve communication performance.
Similarly, RPC libraries such as gRPC~\cite{grpc} and Thrift~\cite{thrift} and communication services such as Kafka~\cite{kafka} and Google PubSub~\cite{gcp-pubsub} implement useful communication primitives. 
In modern deployments, several network functions are often packaged into higher-layer (\ie application-facing) middleware libraries (\eg ServiceRouter~\cite{service-router}) or ``sidecar proxies'' such as Envoy~\cite{envoy} and Istio~\cite{istio}.

The problem, however, is that these communication libraries often make {\em deployment assumptions} to implement their functionality or meet performance requirements.
For example, kernel bypass libraries (and io\_uring under some settings) achieve low latency by requiring that each application thread be ``affinitized'' to a CPU core that is not shared with other threads or processes, thus limiting what other applications the host can run.
Similarly, most pub/sub libraries rely on an external pub/sub service, and pub/sub libraries that use cloud provider services are designed assuming that the application is deployed in the cloud provider's datacenter.

On the other hand, libraries implemented without such deployment assumptions generally perform worse and, in cases such as pub/sub, might not be able to provide equivalent functionality.
Thus, it is appealing (and often necessary) for \emph{developers} to use libraries that make deployment assumptions.

The use of libraries that make deployment assumptions also constrains an application \emph{administrator's} choices, since changing resource allocations or deploying in a different environment can require significant changes to application code. 
For example, reconfiguring a kernel-bypass application to allow sharing application cores with other applications would require changing its code to use a different I/O abstraction (\eg sockets) or switching from polling to interrupt-driven I/O. 
Even when deployed in the same location, an administrator might want to change an application's libraries to cope with changes in workload.
For example, an administrator might want applications to use different pub/sub services (with different cost and scalability trade-offs) depending on the number of subscribers.

For these reasons, when choosing  communication libraries, application developers must consider not only application requirements but also possible deployment locations and workloads.
Accounting for all of these factors, however, is complicated and nearly impossible. Thus, developers today must choose between specialized and performant libraries that limit where an application is deployed and how it responds to workload changes and general libraries that provide flexibility but with poorer performance and fewer features.

This paper presents \name\footnote{Named for Tunnel-Boring Machines.},
a \emph{reconfigurable} and \emph{extensible} network stack that allows developers to use specialized communication libraries and interfaces that make deployment assumptions but enable those decisions to be reconfigured both when deploying an application and at runtime. 
\name does so by allowing application developers to specify several alternate implementations of the same functionality and providing mechanisms to switch between these alternatives.

\begin{figure}
    \centering
    \includegraphics[width=\columnwidth]{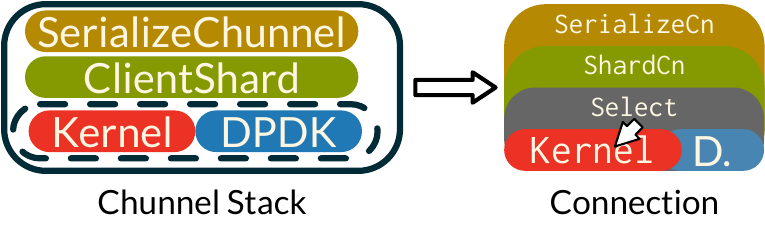}
    \vspace{0pt}
    \caption{A \name connection specifies a \emph{\tunnel stack}, which composes multiple \tunnels. \tunnel stacks can express options (\eg here between Kernel and DPDK), and at runtime \name can create a connection using one set of options (here, Kernel) and also \emph{reconfigure} this choice later (\S\ref{s:reconfig}).}
    \label{f:chunnel-basic}
\end{figure}

\name's core enabling abstraction is the {\em \tunnel}.
A \tunnel, similar to a library, implements some communication functionality, \eg for kernel-bypass networking, serialization, and pub/sub communication.
A developer can implement the \tunnel trait (or interface) to expose some functionality to a \name application.
Applications can \emph{compose} a sequence of \tunnels when creating a connection (Figure~\ref{f:chunnel-basic}).
We use the term ``\tunnel stack'' to refer to the sequence of \tunnels available for a connection; all data sent or received over the connection is processed in sequence by \tunnels in the stack.

Developers specify reconfiguration choices by specifying multiple \tunnels (or a composition of \tunnels) at a particular layer of the stack: in this case, the \name runtime chooses one of these when making a connection and provides a mechanism for the application to change its choice later. We assume that the application can function correctly with any of these \tunnel choices. The only constraint \name imposes on alternate \tunnels in a layer is type safety (\S\ref{s:chunnel-impl}). 
Thus, alternate \tunnels in a layer can make different deployment assumptions and provide different performance and efficiency trade-offs,  allowing application developers to use feature-rich, high-performance communication libraries without limiting an administrator's ability to change deployment environments.

\tunnels enable reconfigurability and extensibility by providing:
\begin{inparaenum}[(a)]
    \item a uniform interface for receiving and sending data, ensuring that switching between different \tunnels does not require any changes to application code or logic;
     \item a typed interface, allowing \name to check during compilation that reconfiguration cannot change the type of input data (\eg bytes, strings, or objects) that a \tunnel processes, nor change the type of outputs it produces; and
     \item embedded runtime type information that allows \name to ensure compatibility between the \tunnels used by a connection's endpoints.
\end{inparaenum}

We have implemented \name's core features in \textasciitilde5,000 lines of Rust. 
In \S\ref{s:applications}, we use an example ETL application implemented using \name to evaluate the performance benefits \name can provide by reconfiguring the network stack and find. \name can unlock application optimizations such as using client-side sharding to achieve $16\times$ higher load meeting a $50$ $\mu$s target latency for a sharded key-value store (\S\ref{s:app:lb}) or 63\% lower latency for a publish-subscribe benchmark (\S\ref{s:app:pubsub}). 
Further, our evaluation in \S\ref{s:eval} shows that \name's abstractions that enable reconfigurability and extensibility add no overhead for most applications and only a 27\% throughput overhead in the worst case (for the highest-performance applications using minimum-sized packets).

\section{Related Work and Motivation}\label{s:relwork}

Reconfigurability and extensibility have been extensively studied in the systems literature. 
In this section, we first outline the prior work on building reconfigurable and extensible systems, then discuss why it is time to make the network stack extensible and reconfigurable.

\paragrapha{Reconfigurability in OSes} Prior approaches to program reconfiguration and extension rely on patching or replacing code in an executable system:  Ksplice~\cite{ksplice} and Proteos~\cite{proteos} developed approaches to patching kernel code to address security concerns, Nooks~\cite{nooks} and Shadow Drivers~\cite{shadow-drivers} describe approaches to replace drivers to avoid failures, and other systems~\cite{wfpatch, pylive} have provided tools for live-patching running applications. Another line of work, exemplified by VM-PHUU~\cite{vm-phu} and NetKernel~\cite{netkernel} develops approaches to reconfigure executing VMs.

\paragrapha{Service Mesh} Service meshes~\cite{service-router, envoy, istio} implement communication functionality outside the applications, making it easier to reconfigure and change features such as how connections are encrypted, or what data is logged. \name allows similar functionality, but unlike these, implements mechanisms to ensure that communicating endpoints have compatible configuration and to allow applications to vary configuration depending on what endpoint they are connected to.

\paragrapha{Reconfigurability in Traditional Network Stacks}
A lot of an application's communication functionality is split across the kernel and the user-space (\eg in communication libraries, service proxies, etc.), and its correctness can depend on agreement between endpoints (\eg agreement on serialization format). Therefore, work on reconfigurable network stacks has largely focused on individual pieces of functionality:
TAPS~\cite{taps} and the Fox project~\cite{biagioni} sought to modularize transport layer functionality.
The x-Kernel's Protocol Framework~\cite{xkernel} proposed an architecture for an extensible network stack, and Microprotocols~\cite{bhatti} built on this work with a proposal to extend the network stack with higher-level protocols such as atomic multicast.
Meanwhile, another early-stage workshop paper~\cite{bertha-hotnets} described a uniform abstraction for using offloads in the network stack, but without reconfiguration.

\paragrapha{Hardware Portability} Since the early 1960s, many systems have aimed to provide \emph{portability}---rather than reconfigurability---thus allowing the system to work on a variety of hardware. Early examples include Unix V4, which was written in C to enable this. Recent examples include DPDK~\cite{dpdk} (which supports multiple NICs), Demikernel~\cite{demikernel} (which supports multiple network technologies including RDMA and kernel bypass networking), and Terraform~\cite{terraform} for configuring infrastructure across different cloud providers. These approaches allow the program to run on multiple deployments, but do not specialize the program to these deployments, and might thus carry a performance or efficiency overhead.

\subsection{What is different now?}
This paper is predicated on the argument that it is now necessary to evaluate and adopt reconfigurable network stacks. This is because of the following three changes that applications must tackle:

\paragrapha{The ``Narrow Waist'' has moved up.} While traditionally applications only required TCP/IP connections that they could establish using the socket API, this is no longer the case~\cite{http-narrow-waist}; modern applications need to adopt one of a variety of open-source libraries that implement more advanced communication abstractions that improve network security, simplify application management, and enable better logging and debugging. These libraries differ in their performance, features, and requirements, and thus, the ability to reconfigure which library is used can significantly reduce management and performance overheads for applications.
For example, applications have several options when choosing how to serialize data, each of which has different tradeoffs:
Apache Arrow is optimized for fast computation, Capnproto can reduce serialization overheads but requires the machine architecture to be identical between communicating hosts, and Protobufs are more general but sacrifice performance. The best choice varies by workload and deployment and thus requires assumptions beyond what the application developer might know.

\paragrapha{The emergence of deployment specific offloads.} Most applications today run in virtualized environments (\eg cloud). These environments increasingly offer specialized services, including hardware offloads (\eg Google's offload-optimized PSP encryption or network sequencers such as Hydra~\cite{hydra}), and managed communication services (\eg publish-subscribe) that can improve application performance. To take advantage of this, applications must explicitly be designed to work in these environments. One approach to doing so is to limit where the application can run or to customize and maintain applications for each runtime environment, but these options decrease utility and increase cost. \name's reconfiguration can embrace environment-specific services without limiting an application's deployment or its network logic.

\paragrapha{Application performance and efficiency matter.} Applications also have more stringent performance and efficiency requirements now, and this limits the ability to use proxies or eschew the use of deployment-specific hardware or software. Reconfiguration is thus useful to meet performance and efficiency requirements without imposing a significant developer burden or limiting where an application is deployed. 

\vspace{0.01in}
Our goal in this paper is to describe a reconfigurable network stack that is useful for today's applications and their diverse runtime environments.
While our approach borrows from prior work---\eg our approach to safe reconfiguration is similar to the one used by Shadow Drivers and Proteos---our main contribution is showing how an extensible abstraction for application networking functionality can enable reconfigurable communication stacks for modern applications.
To build such an abstraction, we addressed three challenges:  (i) ensuring that the component abstraction (we refer to these as \tunnels) can be composed (we refer to these compositions as \tunnel stacks); (ii) ensuring that reconfiguration avoids incompatibility between connected hosts, which we enforce using a negotiation protocol (\S\ref{s:negotiation}); and (iii) minimizing overheads introduced our abstractions, which we do by carefully designing the \tunnel interface and runtime to make use of modern low-cost language abstractions. To the best of our knowledge, no prior work has used these or alternate approaches to provide mechanisms that allow administrators to reconfigure all layers of an application's network stack.

\begin{table}[t]
    \centering
    \small
    \begin{tabular}{p{1.7cm} p{5.7cm}}
        \tunnel & A specific piece of network functionality. \\
        \hline
        \tunnel stack & An application's specification of the set of \tunnels it wants to use. \\
        \hline
        Reconfiguration & Picking or changing \tunnel implementations for a connection at runtime. \\
        \hline
        Negotiation & Ensuring that \tunnel implementations are compatible across connection endpoints. \\
    \end{tabular}
    \vspace{10pt}
    \caption{Glossary of terms used in this paper.}
    \label{t:glossary}
\end{table}

\section{\name Design}\label{sec:design}\label{s:design}
\name is a \emph{reconfigurable} and \emph{extensible} network stack. We start by defining these terms.

\paragrapha{Reconfigurability} refers to an application's ability to choose between different implementations of the sam functionality both \emph{when} establishing a connection and \emph{after} the connection has been established. In practice, this means \name allows applications to choose among \tunnels that implement similar functionality, \eg between DPDK and kernel networking in Figure~\ref{f:chunnel-basic}. As we discussed earlier, choosing \tunnels (\ie reconfiguration) when establishing connections allows applications to be specialized to where they are deployed. Similarly, reconfiguration after a connection has been established is useful when \tunnel performance and efficiency varies depending on workload characteristics or resource availability, since it allows application to respond to changes without disrupting ongong connections. For example, DPDK \tunnels can be built either so they poll for I/O in the application thread---which requires pinning the application thread to a dedicated core but provides low latency---or to use a shared I/O core for polling---which makes it easier to share cores between applications but increases latency due to IPC. The choice between these two alternatives depends on the number of cores that are available and on the observed workload. \name's runtime reconfiguration process allows administrators (or orchestrators) to change how I/O is performed at runtime, without disrupting existing connections.

\paragrapha{Extensible} refers to application and library developer's ability to add new \tunnels to \name. These \tunnels can either represent alternate implementations of the same functionality or entirely new functions. Extensibility allows \name's reconfiguration to allow changes not just to the lowest layers of the network stack, but also to allow changes to how more complex communication primitives are implemented.

Supporting both reconfigurability and extensibility imposes four design requirements on \name:
\begin{compactenum}
\item Switching between two \tunnels that implement the same functionality should not require changes to application logic.
\item Reconfiguration should be \emph{safe}; it should not lead to scenarios where a connected host can no longer communicate. Some \tunnels, \eg serialization \tunnels, require all communicating hosts to use compatible formats, and \name must ensure that this holds.
\item Reconfigurability should neither limit what \tunnels an application can compose nor limit what functionality a \tunnel can implement.
\item When possible, reconfiguring an established connection should not require re-establishing the connection state.
\end{compactenum}

The \tunnel abstraction meets most of these requirements by enforcing a unified interface that ensures that \tunnels can be composed and reconfigured. However, the abstraction alone is insufficient for ensuring safety, and the \name runtime implements a negotiation protocol (\S\ref{s:negotiation}) that checks compatibility between hosts when establishing or re-configuring connections.
For reference, we provide in Table~\ref{t:glossary} a glossary of \name's important concepts.

\begin{listing}[t!]
\begin{minted}[linenos, breaklines]{rust}
// Client-side application code.
let sock = KernelUdpChunnel::new(); // () -> bytes
let shrd = ClientShard::new(cfg); // T -> T + sharding
let ser = SerializeChunnel::new(idl); // bytes -> type T
let conn = bertha::make_stack!(ser, shrd, sock).connect(addr);
\end{minted}
\vspace{-12pt}
\caption{An application developer specifies a \tunnel stack (\texttt{shrd}, \texttt{ser}, and \texttt{sock}). Each of the \tunnels in the stack modify the resulting connection type.}
\label{l:chunnel-stack}
\vspace{-10pt}
\end{listing}
\begin{figure}
    \centering
    \includegraphics[width=\columnwidth]{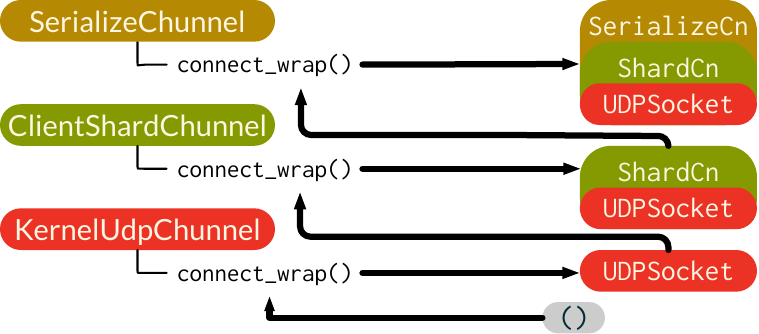}
    \vspace{2pt}
    \caption{\name builds connections by recursively calling \tunnels' \texttt{connect\_wrap} functions. We show reconfiguration in \S\ref{s:reconfig}.}
    \label{f:chunnel}
\end{figure}

\subsection{The \tunnel Abstraction}\label{s:chunnel}
\tunnels are \name's core abstraction: each \tunnel represents a single communication function, \ie logic that can:
\begin{inparaenum}[(a)]
\item Transform data (\eg serializing, encrypting, or compressing data);
\item Decide where to send data (\eg which shard or DHT node should receive a request), or replicate data to multiple endpoints (\eg to implement publish-subscribe functionality); or
\item Send and receive data from hardware (\eg via DPDK or the OS networking stack)
\end{inparaenum}

Each \tunnel takes as input an existing connection $c$, and produces a new connection that encapsulates connection $c$ and adds the \tunnel's functionality to $c$.
Developers of communication libraries implement \tunnels to expose their library's functionality to application developers.
\tunnels are composable: composing two \tunnels creates a third \tunnel.\footnote{\tunnel composition is associative but not commutative.}
By composing \tunnels into \tunnel stacks, application developers can layer functionality into their connections: after applying one \tunnel to a connection, \name can apply another \tunnel to the resulting connection to create a connection that incorporates the features both \tunnels express.
To bootstrap a connection to start with, \tunnels that can appear at the bottom of the stack need to be able to transform a connection with no implementation (represented by a unit type in Rust).

We show an example of using a \tunnel stack in an application in Listing~\ref{l:chunnel-stack}.
Note that we implemented \name in Rust (we discuss this in \S\ref{s:impl}), and we thus describe our interfaces using Rust code. However, the core ideas and abstractions we describe here are general and are readily expressible in other languages.
On line $2$, we show an example of a \tunnel, the \texttt{KernelUdpChunnel}, that transforms the unit type into a UDP connection that can send and receive byte arrays.
On line $3$, we compose this with a \texttt{SerializeChunnel} that transforms this connection into one that exposes an object interface.
Then, on line $4$, we again compose this \tunnel with a sharding \tunnel. This produces a connection that can evaluate a sharding function and rewrite a message's destination address to the appropriate shard.
Finally, on line $5$ we make a \tunnel stack from these three \tunnels, and create a connection with all the specified features for the application.

\begin{listing}[t]
\begin{minted}[linenos, breaklines]{rust}
// A Chunnel control path.
pub struct AChunnel { /*...*/ }
// The ChunnelTransformer<R> trait implements connection establishment logic.
impl<R> ChunnelTransformer<R> for AChunnel
where R: ChunnelDatapath</*...*/>> {
  // Specify that AChunnelDP is the datapath used
  type Connection = AChunnelDP<R>;
  // Chunnel composition interface: compose AChunnel with inner.
  fn connect_wrap(&mut self, inner: R) -> 
    Self::Connection {/*...*/}
  // Specify relative compat. with other impls.
  type Capability = /*..*/;
  fn capabilities() -> Self::Capability { /*...*/ }
}
impl AChunnel {
  // Create a new AChunnel
  pub fn new(/*...*/) -> AChunnel { /* ... */}
}
// The AChunnel datapath
pub struct AChunnelDP<R>{/*...*/}
impl<R> ChunnelDatapath for AChunnelDP<R>
where R: /* input data type requirements */ {
  type Data = /*..*/;
  fn send(&self, ms: impl Iterator<Item=Self::Data>) { /*..*/ }
  fn recv(&self, buf: &mut [Option<Self::Data>]) { /*..*/ }
}
\end{minted}
\vspace{-11pt}
\caption{An overview of the \tunnel interface. Implementors of \texttt{ChunnelDatapath} are connection types, and implementors of \texttt{ChunnelTransformer} allow us to create connection types either from scratch (\ie \texttt{R = ()}) or from an input \texttt{ChunnelDatapath}.} \label{l:chunnel}
\vspace{-10pt}
\end{listing}

\subsection{Implementing a \tunnel}\label{s:chunnel-impl}
To expose their library's functionality as a \tunnel in \name, library developers provide a function that implements the transformer functionality as well as a returned connection type.
We show an example of how a library developer would implement these in Listing~\ref{l:chunnel}.
\name exposes the transformer functionality via the \texttt{ChunnelTransformer} trait (on line 4), and connection-level functionality via the \texttt{ChunnelDatapath} trait (on line 21).
The connection-level \texttt{ChunnelDatapath} trait is straightforward: it defines its data type (line 23) and functions to implement \texttt{send()} and \texttt{recv()}. \texttt{send()} accepts a batch of messages to send, and \texttt{recv()} waits for incoming messages and writes them into a caller-provided array. 

Meanwhile, the \texttt{ChunnelTransformer} trait's function, \texttt{connect\_wrap} (line 9) composes \tunnels in a connection. Figure~\ref{f:chunnel} shows this process. 
If \tunnel A appears above (\ie is pushed after) \tunnel B in a connection's stack, \name calls \tunnel A's \texttt{connect\_wrap} function with an \texttt{inner} argument corresponding to \tunnel B's \texttt{ChunnelDatapath} return type. 
\tunnel A's \texttt{connect\_wrap} function then returns a type that also implements \texttt{ChunnelDatapath}, \texttt{AChunnelDP}, whose data is first processed with \tunnel A's functionality and then passed to \tunnel B (by calling \texttt{ChunnelDatapath} methods on its inner \texttt{R}). 
Each \tunnel's \texttt{connect\_wrap} function is also responsible for performing connection initialization tasks, including initializing connection state, beginning connection handshake, etc. \name calls \texttt{connect\_wrap} recursively (starting with the bottom) on all \tunnels in the connection's stack, thus initializing all \tunnels used by the connection.

Observe that the \texttt{ChunnelTransformer} and \texttt{ChunnelDatapath} take the inner connection's type as a generic type parameter (\texttt{R}).
This allows the compiler to enumerate all possible connection types and optimize them at compile time, thus avoiding most dynamic dispatch overheads.
We discuss how \name encapsulates reconfigurable \tunnels in \S\ref{s:impl}.

Developers can port existing libraries to \name by encapsulating their functionality in \tunnels.
We did this for our implementation, and as we discuss in \S\ref{s:impl}, the largest \tunnel we implemented is a DPDK \tunnel (1,700 lines of Rust code).
Most communication libraries are already designed to be composed with application logic and functionality from other libraries; \name's \tunnel abstraction merely enforces a unified interface that these libraries must implement. This unified interface both simplifies composition and enables reconfiguration. We have implemented several \tunnels and \name applications that we describe in \S\ref{s:applications}, demonstrating the generality of this abstraction.

\paragrapha{\tunnel Datapath Type Safety}
\tunnel stacks are type-safe: we use the Rust type system to ensure that when assembling a connection from a \tunnel stack, the data types of the resulting \texttt{ChunnelDatapath} will be compatible at each layer of encapsulation.
This is because \tunnel implementors specify their data type (\texttt{type Data} on line 23) as well as type restrictions on the connection type their \texttt{ChunnelTransformer} can accept (restrictions on \texttt{R} on line 5, restrictions elided).
\name imposes type restrictions on the input \tunnel stack (implemented as part of the \texttt{connect} function in Listing~\ref{l:chunnel-stack} line 5). So, if one of the \tunnels in a \tunnel stack does not match the others' data types, the application will fail to compile.
For example, a \tunnel that implements sharding needs access to a key that it can use to identify the correct shard. Such a tunnel can specify an input type such as \texttt{(String, Vec<u8>)} to indicate that it requires a string key to be passed along with the data. 
Naturally, datapath type safety does not extend across the network; this is why negotiation (\S\ref{s:negotiation}) is necessary.

\section{Reconfiguring \tunnels}\label{s:reconfig}
\begin{listing}[t]
\begin{minted}[linenos, breaklines]{rust}
// Server-side code to "select" a peer-compatible sharding impl. and between DPDK datapaths. 
let ser = SerializeChunnel::new(/*...*/);
let shrd = select!(MboxShard::new(/*..*/), 
        ClientShard::new(/*..*/));
let dp = select!(DpdkThread::new(/*...*/),
    DpdkInline::new(/*...*/));
let handle = dp.handle();
// conns: a stream of incoming connections.
let conns = bertha::make_stack!(ser, shrd, dp).listen(addr);
/* application logic... */
// trigger a re-configuration to the `DpdkInline` chunnel.
handle.reconfigure(DpdkInline);
\end{minted}
\vspace{-10pt}
\caption{\name applications use \texttt{select} in their \tunnel stack to specify options for reconfiguration.}\label{l:select}
\vspace{-10pt}
\end{listing}
\noindent
The uniform interface \tunnels provide enables reconfiguration in \name.
Recall that each \tunnel in \name is associated with a datapath structure that implements the \texttt{ChunnelDatapath} trait. 
A \name connection that uses the \tunnel instantiates an instance of its datapath structure; the \tunnel's connection state is stored in this structure.
Reconfiguring, \ie replacing one \tunnel with another, thus merely requires transferring connection state from the old \tunnel's datapath structure to the new one, and swapping it into the connection structure. 
We detail our reconfiguration API and mechanism below.

\subsection{Reconfiguration API} 
Listing~\ref{l:select} shows how an application might implement a function that reconfigures connections.
Developers can express a choice between \tunnels using  \name's \texttt{select} type, shown on lines 3 and 5.
The \texttt{select} type encapsulates both \tunnels and uses one of them (according to a developer-specified preference order) when instantiating a connection.
\texttt{select} can also encapsulate two \tunnel \emph{stacks}, which can in turn contain additional instances of \texttt{select}. When instantiating or reconfiguring a connection, \name considers all stacks in the resulting tree of possibilities.
Once a connection is established, the application can use the returned handle (line 7) to call \texttt{reconfigure} and change this decision.

When implementing the \texttt{reconfigure} function, \name must take into account the type of \tunnel being reconfigured. 
All reconfiguration requires changing the \tunnel implementation used by the host;
we refer to this step as \emph{unilateral} reconfiguration (\S\ref{s:impl-reconfig}). 
Reconfiguring some \tunnels not only requires changing implementations but also requires agreement between all communicating hosts. 
We refer to this step as \emph{multilateral} reconfiguration and reuse \name's reconfiguration (\S\ref{s:negotiation}) to ensure safety in this case.
The DPDK \tunnel (line 5--6) is an example of a \tunnel that only requires unilateral reconfiguration, while the sharding \tunnel (line 3--4) requires both unilateral and multilateral reconfiguration.
\tunnel implementors indicate whether multilateral reconfiguration is necessary by implementing a \texttt{capabilities()} function (Listing~\ref{l:chunnel} lines 12-13). 

\begin{figure}
    \centering
    \includegraphics[width=\columnwidth]{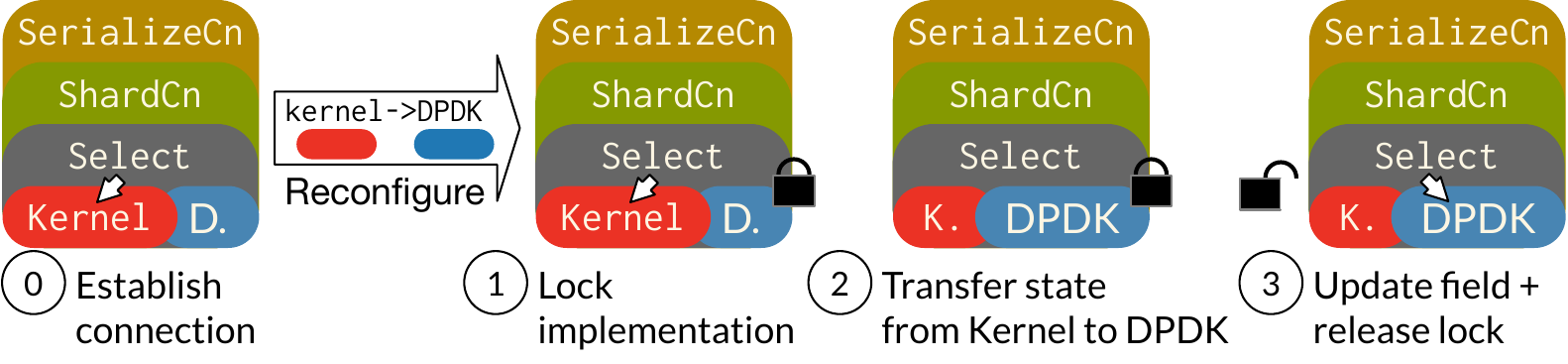}
    \vspace{2pt}
    \caption{\name reconfigures connections using \texttt{Select}, which can safely change its encapsulated \tunnel implementation.}
    \label{f:reconfig}
\end{figure}

We show an example of reconfiguration in Figure~\ref{f:reconfig}, where we replace one \tunnel, a Kernel UDP socket \tunnel, with a DPDK \tunnel in the connection.
Note that multiple \tunnels can be reconfigured at a time, in which case \name amortizes the cost of each step.
First, \name checks compatibility between DPDK's implementation and implementations used by other endpoints in the connection. In this case, the \tunnel is unilateral, so the change is trivially compatible.
Then, \name calls DPDK's \texttt{connect\_wrap} function to initialize it (recall that DPDK's \texttt{connect\_wrap} function bootstraps connections from the unit type). Note that changing this specific \tunnel implementation requires reconfiguring all the connections on the host, since using DPDK often requires exclusive access to the NIC.
This is one reason why the lock shown in Figure~\ref{f:reconfig}'s step \raisebox{.5pt}{\textcircled{\raisebox{-.9pt} {1}}} is necessary; we discuss this further in \S\ref{s:impl-reconfig}. 
Then, \name helps the new DPDK connection type translate the connection state from the kernel's \tunnel implementation (\raisebox{.5pt}{\textcircled{\raisebox{-.9pt} {2}}})
For these \tunnels, the state is simply a list of active connections and the ports they are listening on, but other \tunnels may include other useful state such as sequence numbers.
Finally, \name switches
over to the new implementation (\raisebox{.5pt}{\textcircled{\raisebox{-.9pt} {3}}}).
  
\vspace{-3pt}
\subsection{Reconfiguration Mechanism}\label{s:impl-reconfig}
Here we explain how we switch implementations on a single host.
The core challenge lies in ensuring that \name preserves state as it switches \tunnel implementations. 
Individual \tunnels may have state (\eg the DPDK example from before relies on host-wide NIC configuration), so when re-configuring a connection \name allows the new \tunnel to initialize the corresponding state.
Thus, for safety, we need to ensure that there is a ``switch point'' after which no connection on any thread uses the old implementation, or uses state held by the old implementation.
\name must ensure that the state is copied to the new implementation before this switch point.

We can implement reconfiguration by protecting the encapsulated connection with a lock. The \texttt{reconfigure()} call simply initializes the new encapsulated connection, locks the connection state, and updates it to use the newly initialized connection.
In this implementation, the switching point in this implementation is when \texttt{reconfigure()} releases the connection lock.
We describe a lock-free optimization in \S\ref{s:impl}.

When re-configuring a multilateral \tunnel, we must ensure that all peers agree on when to switch to a new \tunnel stack. 
Therefore, the negotiation protocol occurs while \texttt{reconfigure()} holds the connection lock.
\name listens for negotiation protocol packets concurrently with the application. If it receives one, it takes the connection lock (this is the switching point) before sending any negotiation response that would result in a reconfiguration to ensure there is no race condition between the negotiation response and the subsequent application data packet from the application. 
During reconfiguration, \name uses a two-phase commit to ensure that all endpoints agree on when to switch.
If any peer refuses to switch (or times out), reconfiguration fails.\footnote{Because all peers must accept the transition for it to commit, a faulty peer cannot force other connection participants to switch \tunnel stacks.}
We demonstrate this process in \S\ref{s:app:pubsub}.

\section{Negotiation}\label{s:negotiation}
\name's negotiation protocol ensures compatibility between the \tunnel stacks used by hosts communicating using a connection and is used when connections are established or reconfigured. 
We say two \tunnels are \emph{compatible} if data sent by one can be processed by the other.
For example, two serialization \tunnels are compatible if data serialized by one can be deserialized by the other. 
Checking compatibility requires knowledge of the \tunnel stack being used at all endpoints and thus requires communication.
In this section, we first describe our negotiation protocol for connections that have two endpoints, \ie point-to-point connections.
Then in \S\ref{s:multiparty} we describe a more general protocol that supports connections with more than two endpoints. 
The larger number of endpoints brings additional challenges and complexity, and hence, we use both protocols, choosing between them depending on the type of connection.

\subsection{Negotiation Protocol}\label{s:neg-proto}
To begin negotiation, \name needs a channel over which it can communicate with its peer.
Therefore, we require that the bottom layer of the \tunnel stack always be able to establish a compatible initial connection (by transforming the unit type \texttt{()}). 
We refer to this initial connection as \emph{base connection}, and only require that it provide best-effort delivery.
\name implements a simple reliability and ordering protocol that it uses for negotiation.

Using the base connection, the client sends the server a message describing its \tunnel stack (represented as a set of \tunnel options). 
On receiving this message, the server checks whether its connection has a \tunnel stack (making a choice for each instance of \texttt{select}) that is compatible with the client's options (\S\ref{s:check-compat}). If it finds a compatible stack, it sends it to the client; otherwise, it returns an error to the client and terminates the connection.
Both the server and the client then use recursive \texttt{connect\_wrap} calls (\S\ref{s:chunnel}) to initialize the selected stack.

Each \name connection is associated with a single address, which is used during negotiation.
Therefore, \name does not support connection-less sockets.
This restriction also holds for multi-party connections (\S\ref{s:multiparty}), where the address might represent a pub/sub topic or a multicast IP address.
In cases where a \tunnel, \eg a sharding \tunnel, can send data to multiple endpoints, we require either multi-party negotiation (\S\ref{s:multiparty}) or that all endpoints support the same set of \tunnel implementations.
This ensures that negotiating with one endpoint is sufficient to ensure compatibility with the rest.
Thus, after negotiation, \name returns a nonce that encodes which \texttt{select} branches to use;
a \tunnel can use this nonce to inform endpoints about the stack they should use with the given client.
We adopted this approach for the load-balancing \tunnel in \S\ref{s:app:lb}.

\subsection{Checking Compatibility}\label{s:check-compat}
The negotiation protocol requires \name to check compatibility between pairs of \tunnels. 
A tempting approach to do so would be to use static analysis to check compatibility between \tunnel implementations. However, \name does not constrain how developers write \tunnels, and checking compatibility is at least as hard as checking equivalence, and well-known results in logic show that checking equivalence between programs is undecidable~\cite{sipser13}. 
Therefore, in \name, we adopt a simpler approach where \tunnels provide runtime type information
that \name uses to check compatibility.
However, we allow for a more general version of compatibility and allow \tunnels with different types to be compatible.
To allow for this, \name requires developers to provide capabilities (opaque to \name) that implement a comparison function for checking compatibility between \tunnels (Listing~\ref{l:chunnel} lines 12-13). 

\tunnel developers can either define a new capability type (or reuse an existing one to indicate compatibility with that \tunnel), and return an instance of this type when \name calls the \tunnel's \texttt{capability} function. 
A capability type needs to implement the \texttt{Capability} interface, which requires defining a comparison function, and must be serializable. We do not assume that capabilities are standardized; instead, we only require that developers writing networked programs that communicate with each other (\eg a client and a server) use the same capability types.
Thus, capabilities allow \tunnel developers to indicate \emph{relative} compatibility between different implementations.
For example, if the ProtoBuf developers provide a software reference implementation of their serialization library, a later compatible implementation based on ProtoACC~\cite{protoacc} can reuse the same capability types to indicate compatibility, as can any future implementations.

In practice, we found that we can encode \tunnel capabilities as a set of labels and that checking compatibility between them requires either checking that the sets are equal -- we refer to this as ``exact-match'' -- or that there is a non-empty intersection between sets -- we refer to this as ``composition''.
So, to check \tunnel stack compatibility, \name ensures that ``exact-match'' capabilities (such as serialization) are present in both compared stacks, and that ``compositional'' capabilities (such as sharding) are present in at least one.
If there are multiple possibilities, \name uses the developer's specified preference order (\ie the branch listed first).

\subsection{Multi-Party Negotiation}\label{s:multiparty}
\name supports connections with an arbitrarily large number of endpoints communicating over a single connection (\eg multicast or publish-subscribe). 
Before designing a negotiation protocol for this scenario,  we need to first determine how negotiation works in a multi-endpoint setting. 
A point-to-point connection only exists for as long as the client and server are communicating and does not need to consider cases where endpoints join after the negotiation step has finished,
but with multiple endpoints
we can neither assume that all endpoints are known when a connection is first established nor can we require that all endpoints connect at the same time. 
As a result, multi-party negotiation must allow (a) endpoints to recover a connection's datapath stack even if they did not participate in its negotiation, and (b) when necessary, allow endpoints to trigger another round of negotiation to transition to a different datapath stack. 

We implement a ``rendezvous-based'' negotiation protocol for the multi-endpoint case. We implement negotiation using a key-value store that can be accessed by all endpoints. The key-value store is also responsible for recording the negotiated datapath stack, thus allowing endpoints to recover the connection's datapath stack even when they do not participate in the negotiation. We require that the key-value store support serializable multi-key transactions. However, we do not impose other requirements, and we allow the use of either single-node key-value stores (\eg Redis) or replicated consensus-based stores (\eg etcd). 
While a failure of this key-value store would prevent new endpoints from joining a multi-endpoint connection, it would not impact endpoints that have already joined. 
Furthermore, the negotiation state is not shared across connections, and thus the key-value store can be easily sharded for scalability. While using an external-key value store makes establishing a multi-endpoint connection take longer, any algorithm for multi-endpoint negotiation requires agreement and would thus impose similar performance costs.

A peer starts multi-party negotiation by connecting to the key-value store and proposing a \tunnel implementation stack, using compare-and-swap (\eg via a transaction) to check for an existing stack.
If the compare-and-swap operation succeeds, the peer can safely use the \tunnel implementation stack it proposed. 
Otherwise, the key-value store returns the \tunnel implementation stack already in place amongst the existing connection participants along with the number of participants in the connection.
The new peer then uses the same stack comparison procedure described in \S\ref{s:check-compat} to determine its local stack's compatibility.
The peer can then use this stack to create its connection or propose a reconfiguration as per \S\ref{s:impl-reconfig}.

\section{\name Extensions}\label{s:impl}
We implemented \name in Rust. While our implementation uses several Rust features, including traits, async\footnote{
We implement most \tunnel operations, including \texttt{connect\_wrap()}, \texttt{send()}, and \texttt{recv()}, as async functions. We elide this (as well as Rust's lifetime annotations) in code listings for clarity. We use the popular Tokio~\cite{tokio-rs} async runtime for scheduling the resulting coroutines.
}, and the type system, we believe our ideas are more general and can be implemented using similar features and metaprogramming support in other languages. For example, we believe C++ templates, Racket macros, and Go's generate tool would allow us to implement all of the features we have described in those languages.

We implement \tunnel stacks with a structure similar to Lisp's \texttt{cons}. We represent \texttt{select} simply as a struct that contains both branches. 
Only unilateral \texttt{select}s implement \texttt{connect\_wrap}; \name uses a Rust enum to represent the two choices and returns an instance of this enum.
This enum represents a limited form of dynamic dispatch; at runtime, the connection dispatches to the appropriate implementations of the \texttt{send()} and \texttt{recv()} functions based on the enum's variant.
Multilateral \texttt{select}s cannot directly implement \texttt{connect\_wrap}, since they must use negotiation to pick an implementation branch.
Once the negotiation process returns a choice, the datapath for these multilateral \texttt{select}s is the same as the unilateral case.
We implemented \name's core libraries in \textasciitilde$5,000$ lines of Rust, including the negotiation protocol (of those, \textasciitilde$4,000$ lines).

For the remainder of this section, we describe two optimizations we implemented in \name: zero-rtt negotiation (\S\ref{s:impl:zerortt}), and lock-free single-host reconfiguration (\S\ref{s:impl:fast-reconfig}).

\subsection{Zero-RTT Negotiation}\label{s:impl:zerortt}
As we note in \S\ref{s:negotiation}, runtime-reconfigurability requires agreement, and hence endpoints must communicate before they can transfer data. The point-to-point protocol described in \S\ref{s:negotiation} completes in one RTT. 
To address this, the \name implements an additional optimization that allows clients to reestablish connections without additional negotiation, thus reducing overheads for applications that use many short-lived connections. Our approach to doing so is inspired by QUIC's zero-RTT~\cite{quic} connection establishment.

Zero-RTT negotiation requires the client to remember the datapath stack used by previous connections and re-use it when reconnecting. 
The client sends the server a zero-RTT negotiation message when it knows of a previously negotiated stack and then immediately instantiates that stack; the client can then send data. 
When the server receives a zero-RTT negotiation message, it checks if the previously negotiated stack can still be used. If so, the server re-initiates the stack and uses it to process all subsequent data (including any in-flight data from the client). If, however, the server cannot use the stack, it sends the client a message indicating that the negotiation has failed and proposing a new stack. If \name on the client receives such a message indicating negotiation failure, it tears down the existing stack and instantiates the stack included in the failure message instead.

\subsection{Faster Reconfiguration}\label{s:impl:fast-reconfig}

Changing one or more \tunnel implementations during reconfiguration requires coordinating with any thread that might be using the connection. This coordination is necessary for safety.
The approach we described in \S\ref{s:impl-reconfig} used locks to implement this coordination. While using locks ensures safety, as we show in \S\ref{s:eval:reconfig}, this adds overheads in the common case when no reconfiguration occurs.

We thus also implement a different approach with lower fast-path overheads: during reconfiguration, we wait for all application threads to arrive at a barrier. Once the barrier resolves, threads other than the one performing the reconfiguration wait on a channel before performing any further operations on the encapsulated connection. Thus, we know it is safe to modify the encapsulated connection by simply writing to the appropriate memory\footnote{We use Rust's \texttt{UnsafeCell} to do this.}, since we know there will be no concurrent accesses to the connection object. 
The switching point in this case is the barrier's resolution: after this, no connection will use the old connection objects.
Note that in the case of multilateral \tunnels, the barrier must protect the negotiation phase because the negotiation protocol uses the connection to communicate.
This approach is similar to that used by stop-the-world garbage collectors. The only overhead on the fast path is that when sending or receiving data, each thread needs to read a boolean value indicating whether it needs to stop at a barrier.
We evaluate reconfiguration mechanisms in \S\ref{s:eval:reconfig}.

\cut{
\subsection{\tunnel Stack Optimizations}\label{s:impl:optimization}
The amount of computation performed when sending or receiving data over a connection depends on its \tunnel stack, and the implementations chosen.
In some cases, implementation choices might make redundant some \tunnels in the connection's \tunnel stack, and in other cases reordering \tunnels might result in better performance.
Unfortunately, the specific optimizations that should be applied depend on the connection and on the chosen implementations, which makes it impractical for developers to manually apply these optimizations.
Further, since \name is extensible and cannot know the set of \tunnels it will be working with ahead of time, it cannot implement these optimizations automatically.

\name instead takes an approach inspired by Broadway~\cite{broadway} 
where \tunnel implementors and application developers can provide \emph{optimization passes}.
These passes work on subsets of the \tunnel stack, and can modify them. These passes are similar to optimization passes in machine learning frameworks~\cite{tensorflow-xla, onnx, tvm}, data analytics~\cite{weld}, and Click~\cite{click-opt}.
Passes are used at compile time: \name enumerates all possible \tunnel stacks that can be used by a connection and applies optimizations to them. This is because the passes themselves rely on type information (to identify \tunnel and \tunnel implementations), and this information is not available at runtime.
We believe we can implement runtime optimizations instead by recording type information (similar to RTTI), but we leave this to future work.

Note that \name assumes that optimization passes are correct.
While the type system does check that substitutions are compatible (\eg disallowing substitutions that change the input or output datatype), it can neither ensure that semantics remain the same nor reason about performance.
Note that negotiation will still check the post-optimization \tunnel stack for multilateral compatibility.
Therefore, we require application developers to analyze these effects, and then explicitly opt into optimization passes.
Thus, the effect here is to reduce the tedium and effort required to optimize an application. 
}

\newcommand{\etlapp}{ETLApp\xspace}
\section{Application Benefits Evaluation}\label{s:applications}
In this section, we discuss implementing an ETL (extract, transform, and load) application in \name, and use it to demonstrate how \name can lower the operating cost of applications deployed in the cloud and improve their throughput and latency. We delay an evaluation of \name's abstractions to \S\ref{s:eval}.

\etlapp is a \name based application that processes streaming logs in near-realtime to produce summary statistics and metrics that can be used by users or displayed on a dashboard. \etlapp consists of four microservices:
(1) \emph{producers}, which run at the data source (\ie where logs are generated) and send logs and other telemetry information to (2) \emph{ingesters}, which perform light-weight analysis on these logs to determine how they should be ordered and then publish them to an appropriate topic in a publish-subscribe service. (3) \emph{Parsers} subscribe to these topics, parse log entries published by the ingesters, compute summary statistics and forward them to (4) \emph{consumers}, which store them and allow users or applications to query and retrieve computed statistics.

\etlapp uses a variety of communication functions and \tunnels: the producers use QUIC or TLS (and PSP~\cite{psp}, where available) to protect their communication with the ingester (\S\ref{s:app:quic}) and a load balancer \tunnel to decide what ingester the producer should send data to (\S\ref{s:app:lb}; ingesters and parsers use a Kafka~\cite{kafka} or cloud provider publish-subscribe \tunnel (\S\ref{s:app:pubsub}; and the parser uses a local communication \tunnel that minimizes encryption and transport costs (using techniques inspired by Slim~\cite{slim}) when the parser and consumer microservices are colocated on the same server.

\etlapp consists of \textasciitilde 2,300 lines of Rust code.
The \tunnels we used, which include ones implementing QUIC, TCP, UDP (via the kernel, Shenango~\cite{shenango}, and a custom DPDK datapath), TLS, ordering, four publish-subscribe services, reliability, routing, serialization, and sharding, consist of  \textasciitilde 15,000 lines of Rust code. Individual \tunnels range from \textasciitilde 400 lines for TLS and QUIC \tunnels to \textasciitilde 5,000 lines for a DPDK \tunnel.
\begin{figure}
    \centering
\begin{knitrout}
\definecolor{shadecolor}{rgb}{0.969, 0.969, 0.969}\color{fgcolor}
\includegraphics[width=\maxwidth]{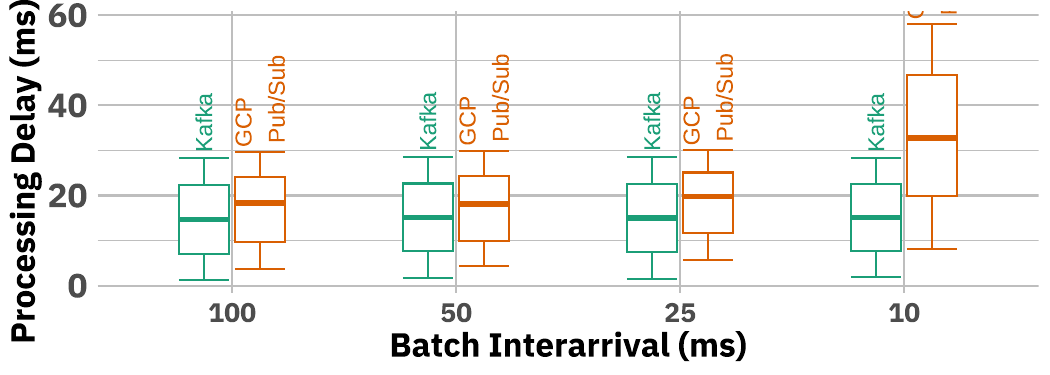} 
\end{knitrout}
    \caption{The ingesters and producers negotiate whether to use Kafka or GCP Pub/Sub. Here, we force the connection to take either choice, respectively, to demonstrate the performance tradeoffs involved.} \label{f:elk-kafka-gcp}
\vspace{-10pt}
\end{figure}

In the remainder of this section, we first evaluate \etlapp's end-to-end performance and demonstrate reconfiguration's performance benefits. After that, we describe reconfiguration in more detail for three cases: QUIC and TLS (\S\ref{s:app:quic}), publish-subscribe message queues (\S\ref{s:app:pubsub} and load balancing (\S\ref{s:app:lb}). 

\paragrapha{End-to-End Performance} We use processing latency as the main metric for evaluating \etlapp's performance. We define processing latency as the time taken between when log messages are available to the producer and when the messages are incorporated in the consumer's summary statistics. Our performance evaluation uses a fixed offered load, so processing latency accounts for both computation time and queueing.
Figure~\ref{f:elk-kafka-gcp} shows \etlapp's processing latency in two configurations: one with a Kafka~\cite{kafka} instance available in the local cluster and one without. 
Deciding whether to run a local Kafka instance or use a managed publish-subscribe service is a common choice application operators must make: running a self-hosted Kafka instance requires constant hardware resources but can be closer to other application components and tuned for the application's workload, offering better performance. Meanwhile, managed services such as GCP Pub/Sub offer a per-message cost model (see \S\ref{s:app:pubsub}) but can have lower performance. \name allows a single \etlapp implementation to work in both cases, since programmers \texttt{Select} between these two alternatives when establishing connections at the ingester and parser. We ran this evaluation on Cloudlab \texttt{m510}
machines using Linux kernel 5.15 with Microk8s version 1.27.5 as a container runtime and Kafka version 3.1.0, and used 8 ingesters and 4 parsers. Every interarrival duration, our workload adds a batch of 16 150 byte records to the producer's log for processing. Our results (Figure~\ref{f:elk-kafka-gcp}) show that using a Kafka \tunnel enables lower processing latency for \etlapp at higher offered loads (or lower interarrival times). 
However, at lower offered loads using the GCP Pub/Sub \tunnel both matches the Kafka configuration's performance and has a lower cost.  As a result, no one configuration is optimal for \etlapp across all scenarios we evaluate.  Indeed, we find similar tradeoffs exist in the other \etlapp components, and we discuss them in the rest of this section.

\subsection{Encryption Options}\label{s:app:quic}
Following standard guidance, we want to ensure that communications between \etlapp's producer and ingester are encrypted. Application developers have several options available to ensure this: they can use HTTP/2 with TLS and TCP, HTTP/3~\cite{http3-rfc} which uses QUIC~\cite{quic} and UDP, academic proposals such as TCPLS~\cite{tcpls}, or cloud-specific hardware encryption such as Google's PSP~\cite{psp} and Microsoft's hardware accelerated TLS~\cite{microsoft-encryption}. Each of these carries different overheads (\eg PSP and other hardware accelerated variants require nearly no CPU cycles, while other options must use CPU cycles for encryption), imposes different deployment limitations (\eg requiring deployment in Google's cloud), and also affects what services an application can access (one cannot use QUIC to access a TLS-only service). 
While there have been ad-hoc efforts to resolve some of these problems, \eg  QUIC servers can advertise QUIC support to TLS clients using an \texttt{Alt-Svc} HTTP header, and TCPLS clients use a TLS extension to similar effect, these efforts are not widely accessible.

\name offers an alternative that avoids deployment and connectivity concerns while using applications to use the most efficient encryption protocol for each connection. This is because \name uses negotiation to agree on what \tunnel to use, thus ensuring compatibility and portability.
In \etlapp, we express these options simply as follows: \texttt{Select(EnsurePSP, Select(QUIC, make\_stack!(TLS,TCP))).connect(..)}.

\eat{
\subsection{Ordering Primitives}\label{s:app:ordering}

A longstanding debate in the systems community considers whether the network abstraction should provide (total or causal) ordering as in ISIS~\cite{virtual-synchrony}, or whether the network should be kept simple and application-level protocols should be used to implement consisted delivery instead.
While application-level approaches have been dominant for some time, recent trends show that we are moving back towards network abstractions that implement ordering~\cite{mom, hydra}.
These new abstractions avoid the pitfalls of earlier implementations by using offload hardware, but as a result they cannot be used in all deployments.
Thus, the best implementation to use depends on where applications are deployed. An ordered delivery \tunnel could mediate the trade-off space between these options: if a network sequencer is available, the connection could use it to offload ordering. Otherwise, it could use traditional software-only consistency techniques.
}

\subsection{Publish-Subscribe Message Queues}\label{s:app:pubsub}
\etlapp uses ordered publish-subscribe (pub/sub) services for communication between ingesters and parsers to ease scaling up ingesters and parsers and simplify fault tolerance. 
Pub/sub services are widely available both as managed cloud services, \eg AWS SQS~\cite{sqs}, GCP PubSub~\cite{gcp-pubsub}, Azure Storage Queues~\cite{az-storage-queues}, and Azure Service Bus~\cite{az-service-bus}), and as open-source software, \eg Kafka, that applications can deploy and manage. We demonstrated above (Figure~\ref{f:elk-kafka-gcp}) that Kafka, when deployed by an application, offers \etlapp better performance at a high load but can incur increased costs. In this subsection, we dive deeper into how the choice of pub/sub implementation can affect ordering and delivery semantics and costs and why workloads and requirements might change which implementation an application uses.

Ordering and delivery semantics can vary significantly across pub/sub services: AWS SQS and GCP PubSub support in-order message delivery, Azure Storage Queues does not guarantee any message ordering, and Kafka always guarantees message ordering. Similarly, in terms of delivery semantics 
GCP PubSub and Kafka allow application developers to create broadcast topics or subscriptions where multiple receivers can receive the same message, but AWS SQS requires the use of a separate service (AWS SNS~\cite{sns}), and Azure Queues does not natively support multicast. Services also exhibit significant pricing variance: in 2023, AWS SQS provides $1$ million free 64KB messages per month and charges per million subsequent unordered (\$0.40) and ordered (\$$0.50$) messages.  GCP PubSub provides $10$GB of free messages per month with a minimum message billing size of $1,000$ bytes, and then \$$40$/TB. Azure Storage Queues~\cite{az-storage-queues} charges \$$0.004$/$10,000$ messages.

\newcommand{\awsImprovementPct}{63}


\paragrapha{Reconfiguring Ordering Implementations}  
As a result, the best option for \etlapp depends on its input workload, since this determines the complexity of application logic required to reorder messages. 
Specifically, we found that when a single parser is in use,  using SQS's best-effort queueing setting and ordering messages at the parser yields \awsImprovementPct\% lower median latency than using a service-ordered message queue. However, this does not work when multiple parsers are in use, since ordering would require coordination across all parsers. 
\name allows \etlapp to switch between both modes: we specify that when using SQS, \etlapp should use ordered queues only when multiple receivers are active, and we use \name's multi-party negotiation to dynamically reconfigure the pub/sub connection.

\begin{figure}
    \centering
\begin{knitrout}
\definecolor{shadecolor}{rgb}{0.969, 0.969, 0.969}\color{fgcolor}
\includegraphics[width=\maxwidth]{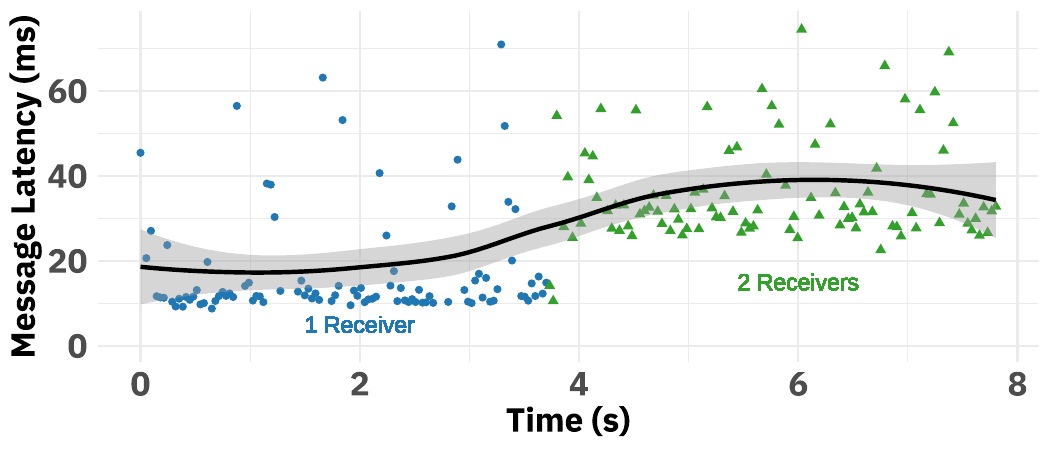} 
\end{knitrout}
    \caption{Runtime reconfiguration allows switching datapath stacks. Note that the first few requests after the second receiver arrives still use the original stack while the transition commits.} \label{f:msgqueue-transition}
    \vspace{-5pt}
\end{figure}

We demonstrate this performance benefit with a microbenchmark: in Figure~\ref{f:msgqueue-transition}, we send $100$ messages with an inter-arrival time of $25$ms, then start a second receiver and send another $100$ messages. We use $5$ ordering groups and AWS SQS. We observe that the application uses receive-side ordering and observes lower message latencies when a single receiver is present and then safely transitions to the higher-latency and more costly service-provided ordering when the second receiver arrives.

\begin{figure}
    \centering
\begin{knitrout}
\definecolor{shadecolor}{rgb}{0.969, 0.969, 0.969}\color{fgcolor}
\includegraphics[width=\maxwidth]{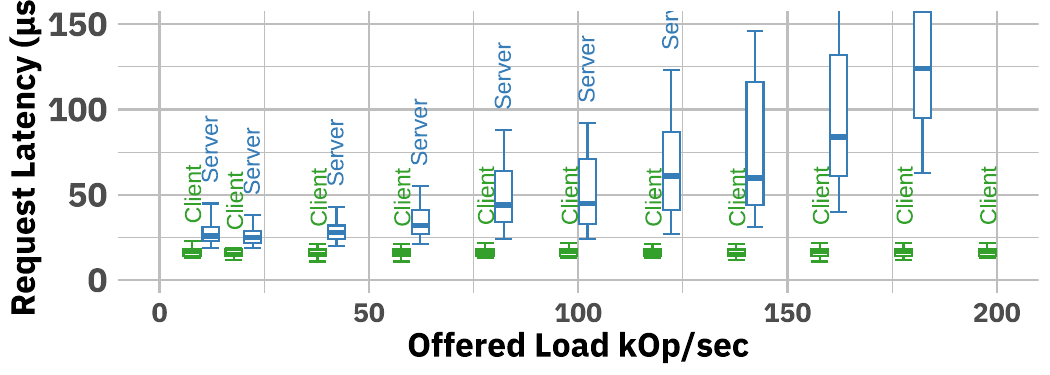} 
\end{knitrout}
    \caption{Client sharding scales better with lower latency, and can take advantage of more concurrency at the client without burdening the server implementation with more connections.} \label{f:kv-client}
\vspace{-10pt}
\end{figure}

\subsection{Load Balancing}\label{s:app:lb}
We use load balancers to spread load between active ingesters in an \etlapp deployment. This allows us to handle cases where the rate of log updates varies across producers and cases where ingesters might straggle. Modern applications commonly employ load balancing for this purpose, and a variety of different load-balancers are available: both cloud managed versions, \eg Amazon's ``Application Load Balancing'' (ALB), Azure's ``Application Gateway'', or GCP's ``Internal HTTP(S) Load Balancing'', open-source versions, and even ones that are incorporated within the application~\cite{service-router}. 
The choice of load balancers has an impact on cost and performance, and the appropriateness of each varies by workload.

Evaluating the benefit of reconfigurable load-balancing using \etlapp is challenging since external factors (\eg where the load balancer is deployed) often determine performance. Therefore, we instead demonstrate this benefit using a key-value store implemented with \name. We also used two load-balancing \tunnels that provide us with greater control and an easier way to reason about performance: one in which a remote server evaluates a hash function on message keys and then forwards messages to the appropriate server, and one where the client evaluates the hash function and sends messages that bypass the load balancer. We evaluated these options using a YCSB-based benchmark. In each experiment, we first run a warm-up phase and loading phase that issues $12,000$ PUT requests, then we use 3 clients, each running on its own server and splitting its requests among $8$ connections, to issue requests according to the ``Workload B'' request distribution.  In this workload, each message is $132$B in size. Each client issues 1,214,020 requests, and we control the request rate (using Poisson arrivals) to meet a target offered load. To ensure that the offered load remains consistent, we terminate the experiment after any connection sends its last request. We report results from evaluation run on  \texttt{xl170} nodes in Cloudlab, which uses 25 Gbit/s Mellanox CX-4 Lx NICs, with one server machine and two client machines. All machines used Linux kernel 5.4.0 and Mellanox OFED driver version \texttt{5.6-2.0.9.0}. We used a custom DPDK-based UDP \tunnel with this application, which uses DPDK version \texttt{20.11.0}. 

Our results in Figure~\ref{f:kv-client} show that client-side load balancing offers better performance: when using server-side sharding, the key-value store cannot meet a $50\mu$s bound for p$95$ latency beyond 40,000 requests per second, but with client-side sharding can meet the latency bound up to 800,000 requests per second. However, client-side load balancing makes application management more complex since changing the set of backend servers requires updating all clients. \name allows application deployers to switch between these options: using the client-side approach in cases where the set of backend servers is relatively stable and using server-side load balancing otherwise.

\section{\name Overheads Evaluation}\label{s:eval}
We next evaluate \name's overheads. We focus on cases that require the most processing, and consequently, we report worse-case overheads for \name. Our evaluation focuses on three aspects of \name's performance: (a) the impact \tunnels have on application throughput (\S\ref{s:eval:tput}); (b) how does \name impact application latency (\S\ref{s:eval:kv-overhead}); and (c) how long \name takes to reconfigure a connection (\S\ref{s:eval:reconfig}).

\subsection{Throughput}\label{s:eval:tput}
\begin{figure}
    \centering
    \begin{subfigure}[t]{\columnwidth}
\begin{knitrout}
\definecolor{shadecolor}{rgb}{0.969, 0.969, 0.969}\color{fgcolor}
\includegraphics[width=\maxwidth]{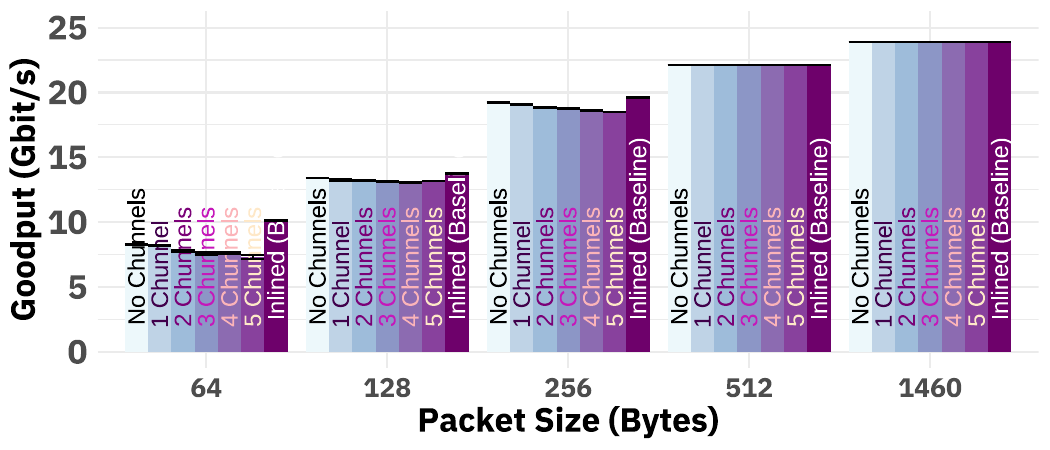} 
\end{knitrout}
        \caption{GBit/s}\label{f:throughput-ubench:bps}
    \end{subfigure}
    \begin{subfigure}[t]{\columnwidth}
\begin{knitrout}
\definecolor{shadecolor}{rgb}{0.969, 0.969, 0.969}\color{fgcolor}
        \vspace{10pt}
\includegraphics[width=\maxwidth]{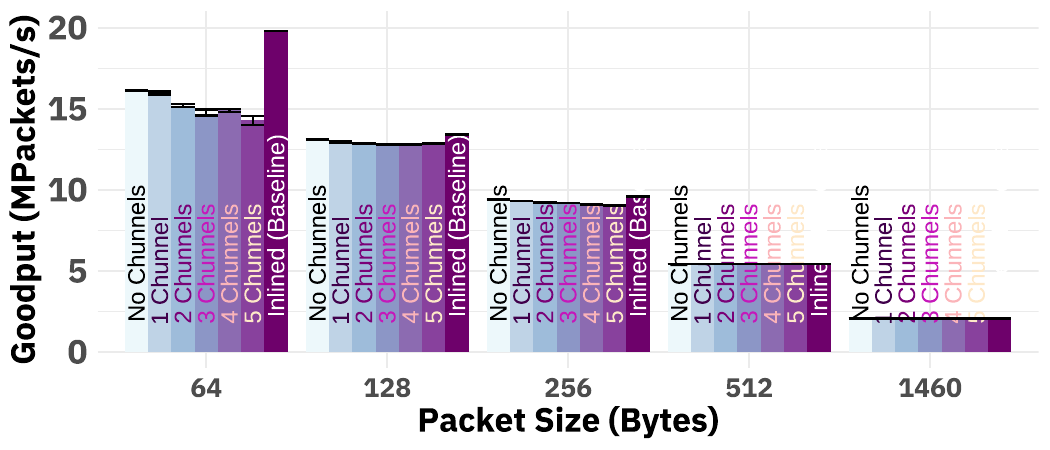} 
\end{knitrout}
\caption{MPackets/s}\label{f:throughput-ubench:pps}
    \end{subfigure}
    \vspace{20pt}
    \caption{The overhead of including a marginal \tunnel in a DPDK-based UDP connection is between 0-27\%, and becomes negligible with larger packet sizes. Columns show p50 and errorbars show p25 and p75 across 10 trials.} \label{f:throughput-ubench}
\end{figure}

\begin{figure}[t]
    \centering
\begin{knitrout}
\definecolor{shadecolor}{rgb}{0.969, 0.969, 0.969}\color{fgcolor}
\includegraphics[width=\maxwidth]{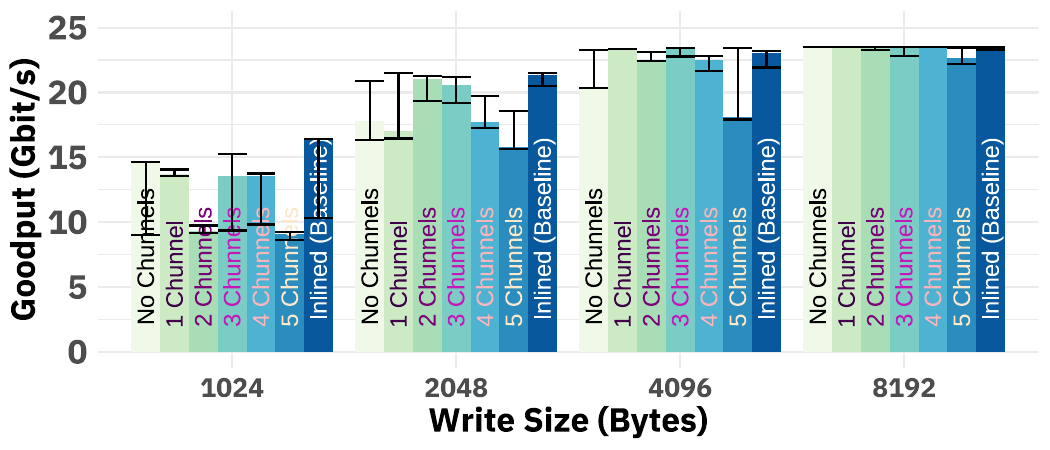} 
\end{knitrout}
    \caption{Overhead of a marginal \tunnel for Linux TCP-based connections.} \label{f:tcp-throughput}
\end{figure}

We evaluate \name's impact on application throughput using a file transfer benchmark where a client uses $10$ threads, each of which connects to a server, sends a short request receives a $5$GB file in response (thus transfering a total of $50$GB. We implemented two versions of this benchmark: a UDP version that uses a UDP \tunnel over a DPDK \tunnel that implements logic for multiplexing and demultiplexing connections, and a TCP version that uses a TCP \tunnel implemented using sockets. We refer to these two base implementations as `No \tunnel' implementation, since each only includes code that any such client would require. As a baseline, we compare this version's throughput to that of a version of the application written without DPDK, which our figures refer to as the `Inlined (Baseline)' configuration.
We also compare the throughput of these `No \tunnel' versions to the throughput achieved by an extension of the program that adds between one and five no-op \tunnels\footnote{We used Rust's \texttt{std::hint::black\_box} in the no-op \tunnel, forcing the compiler to assume that the caller can use any part of the data.} (labelled `1 Chunnel', etc.), that merely reads the data and forwards it to the next \tunnel. 

We measure goodput (\ie rate of file transfer) for both versions of the application, while running them on XL710 nodes connected by $25$Gbit/s links in CloudLab (the same setup as in \S\ref{s:app:lb}). For the UDP applications, we vary packet sizes between $64$ and $1,460$ bytes, while for the TCP applications, we vary the size of the buffer provided to the \texttt{send} call to be between $1,024$ to $8,192$ bytes.

We show the UDP results in Figure~\ref{f:throughput-ubench} and the TCP results in Figure~\ref{f:tcp-throughput}. For the UDP configuration, we find that in the worst case (the $64$Byte packets), the median achieved goodput and packets transmitted are $27\%$ lower with 5 Chunnels and $18\%$ lower with 1 Chunnel. However, even a small increase in packet size to $128$ Bytes decreases this overhead to $4\%$ with 5 Chunnels, and further increases in packet size cause this to decrease even further.
For the TCP configuration, trends are less clear because of the inherent variability in using the kernel networking stack, but as a general trend, the overhead decreases as the write size increases.

\begin{figure}
    \centering
\begin{knitrout}
\definecolor{shadecolor}{rgb}{0.969, 0.969, 0.969}\color{fgcolor}
\includegraphics[width=\maxwidth]{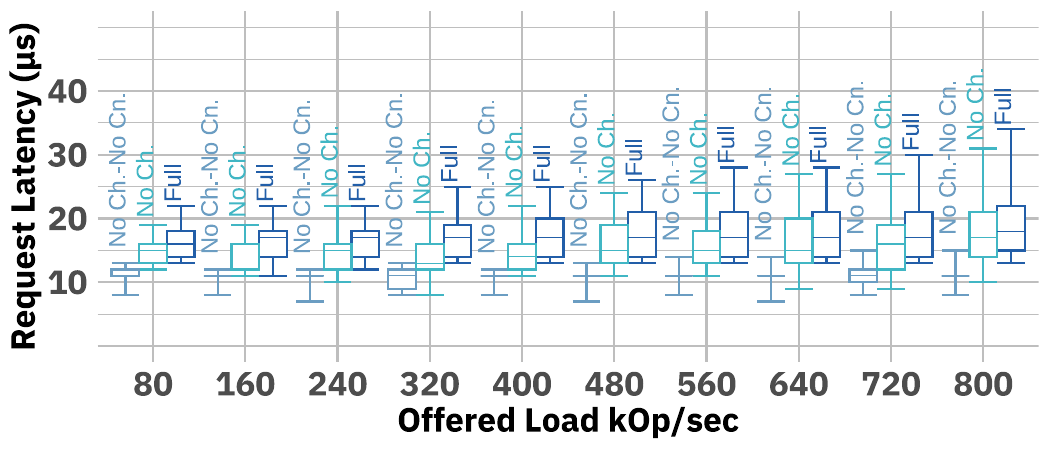} 
\end{knitrout}
\caption{``Full'' shows the KV-store application using a multi-threaded DPDK datapath. ``No Ch.'' removes the use of Chunnels and negotiation, and ``No Ch.-No Cn.'' further removes connection multiplexing and demultiplexing.} \label{f:kv-datapaths}
\end{figure}
\subsection{Key-Value Store Operation Latency}\label{s:eval:kv-overhead}
Next, we use the key-value store benchmark described in \S\ref{s:app:lb} to evaluate \name's impact on latency. In this case we use three functional \tunnels (serialization, sharding, and reliability) and compare \name's latency (referred to as `Full') in our graphs to two baselines: one that removes negotiation and \tunnels from the application and inlines the remaining implementation, but retains connection multiplexing and demultiplexing features in the datapath (referred to as `No Ch.'); and a second that also removes connection multiplexing (`No Ch.-No Cn.').

Figure~\ref{f:kv-datapaths} shows response latencies as we vary the offered load. Compared to `No Ch.-No Cn.', \name (`Full') exhibits $28\%$ higher median latency at an offered load of $80,000$ requests per second, and this $64\%$ higher median latency at $800,000$ requests per second. We analyzed this result further and found that this difference is because of time spent in connection multiplexing and demultiplexing. Consequently,  when we compare \name's latency to `No Ch.', we observe that \name's median latency is within $6\%$ at both $80,000$ and $800,000$ requests per second of offered load. Connection multiplexing and demultiplexing are common requirements for most applications, and therefore, our results show that \name imposes minimal latency overheads while providing additional functionality in the form of reconfiguration.

\begin{figure}
    \centering
\begin{knitrout}
\definecolor{shadecolor}{rgb}{0.969, 0.969, 0.969}\color{fgcolor}
\includegraphics[width=\maxwidth]{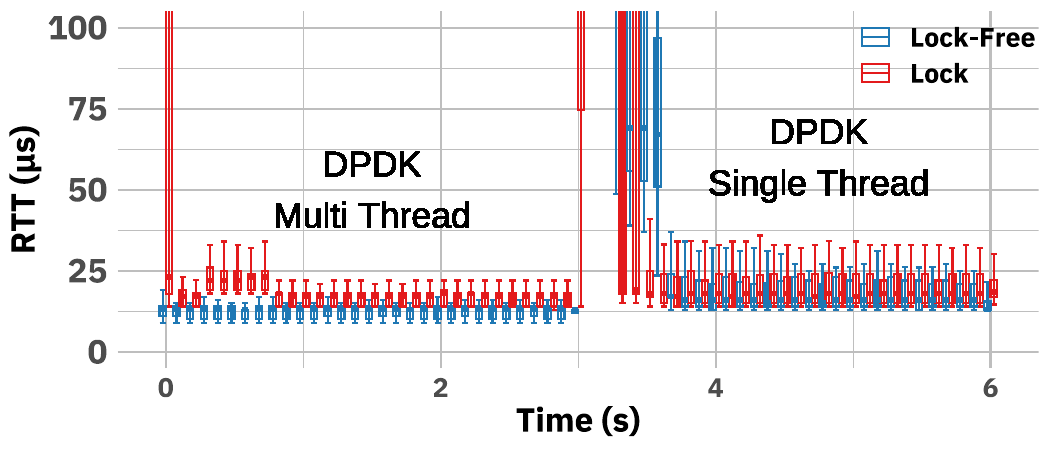} 
\end{knitrout}
    \caption{\name can swap a connection's datapath at runtime without disrupting application state. In this example, we change a server from a multi-threaded DPDK datapath to one that spin-polls only on one thread.} \label{f:dp-swap}
\vspace{-10pt}
\end{figure}
\newcommand{\beforeSwapDpMedianRttUs} {13}
\newcommand{\afterSwapDpMedianRttUs}  {16}
\newcommand{\duringSwapDpMedianRttUs} {68}
\newcommand{\beforeSwapMuxMedianRttUs}{17}
\newcommand{\afterSwapMuxMedianRttUs} {18}
\newcommand{\duringSwapMuxMedianRttUs}{18}

\subsection{Reconfiguration}\label{s:eval:reconfig}
Finally, we evaluate the effect of dynamic reconfiguration on application latency.
Recall from \S\ref{s:impl:fast-reconfig} that \name offers two mechanisms for dynamic reconfiguration, one using a lock and another which is lock-free; we evaluate both implementations.
We implement a microbenchmark that switches between two implementations of a DPDK datapath: ``DPDK Single Thread'', which uses only one spin-polling thread to perform packet operations (similar to Shenango~\cite{shenango} or Snap~\cite{snap}), and ``DPDK Multi Thread'' uses spin-polling coroutines on all threads to perform packet operations.

We show the results of switching from ``DPDK-Inline'' to ``DPDK-Thread'' in Figure~\ref{f:dp-swap}. The echo benchmark triggers a \tunnel switch at the server after 3,000 messages, and we display percentile boxplots (p5, p25, p50, p75, p95) of the achieved latency for the two configurations within each 100ms measurement epoch.
Since the lock-full reconfiguration implementation must acquire a lock for every send and receive, the lock-free version achieves lower latencies with the multi-thread datapath: \beforeSwapDpMedianRttUs $\mu$s median latency vs \beforeSwapMuxMedianRttUs $\mu$s for lock-ful.
However, the lock-free version incurs a larger reconfiguration time: we observe a \duringSwapDpMedianRttUs $\mu$s median latency in the 10 seconds following the reconfiguration vs. \duringSwapMuxMedianRttUs $\mu$s when using a lock. Note that reconfiguration in this case requires reconfiguring the NIC, and this results in a short duration when the application can neither send nor receive network traffic.

\vspace{-3pt}
\section{Discussion and Conclusion}\label{s:discussion}
\paragrapha{Porting Existing Libraries to \name}
To use \name, application developers must find \tunnels that implement the functionality they require.
At present, this might require them to port existing libraries to \name.
Porting most libraries requires minimal effort (as discussed in \S\ref{s:impl}): the developer calls the appropriate library functions from the \tunnel's implementation.
However, building \tunnels is more complicated in two cases: the first is when trying to encapsulate systems, \eg Shenango~\cite{shenango,caladan} and Shinjuku~\cite{shinjuku}, that provide specialized runtimes rather than composable libraries; the second is when dealing with libraries like OpenSSL~\cite{openssl} which limit how they can be composed (OpenSSL uses sockets internally). 
While we found that we could always encapsulate these runtimes, doing so required taking approaches such as running them in a different thread or process and communicating with them using IPC or memory buffers, adding to \name's overhead.

\paragrapha{Deploying \name} 
Because \name introduces a negotiation protocol, all parties in a connection must use it to be able to communicate.
This negotiation protocol can replace existing ad-hoc methods of feature discovery, such as the QUIC discovery convention described in \S\ref{s:app:quic}.
Of course, to use \name's negotiation an endpoint must be aware of it; past work has described various mechanisms for advertising protocol availability such as via DNS, and any such mechanism would work with \name.
While we acknowledge that using \name between arbitrary Internet hosts may be challenging, there are common cases where a single entity controls all connection participants, such as in service proxies,
a cloud provider's SDK, or for internal communication between components of a distributed system.

\paragrapha{Conclusion}
\name provides abstractions that allow the network stack to be reconfigured at runtime.
These abstractions, which provide a uniform interface, not only enable reconfiguration but also make it easy to extend \name with new functionality and allow the \name runtime to ensure check compatibility between communicating hosts, thus improving reliability. We will open-source our implementations and experiment scripts.

\eat{
\name brings reconfigurability to the network stack by allowing applications to defer the decision of which implementations of their communication functionality to use to runtime. 
We show that \name's abstractions also enable extensibility by making it easy to add new functions, improve safety by allowing \name to check compatibility when establishing or reconfiguring connections, and come with small overheads.
}

\label{p:end}
\newpage
\bibliographystyle{abbrv} 
\bibliography{bertha-main}
\end{document}